 \documentclass[preprint,journal]{vgtc}       % preprint (journal style)

\ifpdf%                                % if we use pdflatex
  \pdfoutput=1\relax                   % create PDFs from pdfLaTeX
  \usepackage{graphicx}                % allow us to embed graphics files
  \DeclareGraphicsExtensions{.pdf,.png,.jpg,.jpeg} % for pdflatex we expect .pdf, .png, or .jpg files
\else%                                 % else we use pure latex
  \usepackage{graphicx}                % allow us to embed graphics files
  \DeclareGraphicsExtensions{.eps}     % for pure latex we expect eps files
\fi%

\graphicspath{{figures/}{pictures/}{images/}{./}} % where to search for the images

\usepackage{microtype}                 % use micro-typography (slightly more compact, better to read)
\PassOptionsToPackage{warn}{textcomp}  % to address font issues with \textrightarrow
\usepackage{textcomp}                  % use better special symbols
\usepackage{mathptmx}                  % use matching math font
\usepackage{mathtools}
\usepackage{times}                     % we use Times as the main font
\usepackage{cite}                      % needed to automatically sort the references
\usepackage{tabu}                      % only used for the table example
\usepackage{booktabs}  
\usepackage{amsmath}
\usepackage[dvipsnames]{xcolor}
\onlineid{0}

\vgtccategory{Research}
\vgtcpapertype{application}

\title{THALIS: Human-Machine Analysis of\\ Longitudinal Symptoms in Cancer Therapy}

\author{Carla Floricel, Nafiul Nipu, Mikayla Biggs, Andrew Wentzel, Guadalupe Canahuate,\\ Lisanne Van Dijk, Abdallah Mohamed, C.David Fuller, and G.Elisabeta Marai}
\authorfooter{
\item
C. Floricel, N. Nipu, A. Wentzel, and G.E. Marai are at the University of Illinois at Chicago. E-mail: cflori3@uic.edu.
\item
M. Biggs and G. Canahuate are at the University of Iowa. 
 \item 
L.V. Dijk, A. Mohamed, C.D. Fuller are with the MD Anderson
Cancer Center at the University of Texas.

}

\shortauthortitle{Floricel \MakeLowercase{\textit{et al.}}: THALIS: Longitudinal Symptoms}
\abstract{Although cancer patients survive years after oncologic therapy, they are plagued with long-lasting or permanent residual symptoms, whose severity, rate of development, and resolution after treatment vary largely between survivors. The analysis and interpretation of symptoms is complicated by their partial co-occurrence, variability across populations and across time, and, in the case of cancers that use radiotherapy, by further symptom dependency on the tumor location and prescribed treatment. We describe THALIS, an environment for visual analysis and knowledge discovery from cancer therapy symptom data, developed in close collaboration with oncology experts. Our approach leverages unsupervised machine learning methodology over cohorts of patients, and, in conjunction with custom visual encodings and interactions, provides context for new patients based on patients with similar diagnostic features and symptom evolution. We evaluate this approach on data collected from a cohort of head and neck cancer patients. Feedback from our clinician collaborators indicates that THALIS supports knowledge discovery beyond the limits of machines or humans alone, and that it serves as a valuable tool in both the clinic and symptom research.%
} % end of abstract


\CCScatlist{ % not used in journal version
 \CCScat{K.6.1}{Management of Computing and Information Systems}%
{Project and People Management}{Life Cycle};
 \CCScat{K.7.m}{The Computing Professio
 n}{Miscellaneous}{Ethics}
}

\teaser{
  \centering
\includegraphics[width=\linewidth]{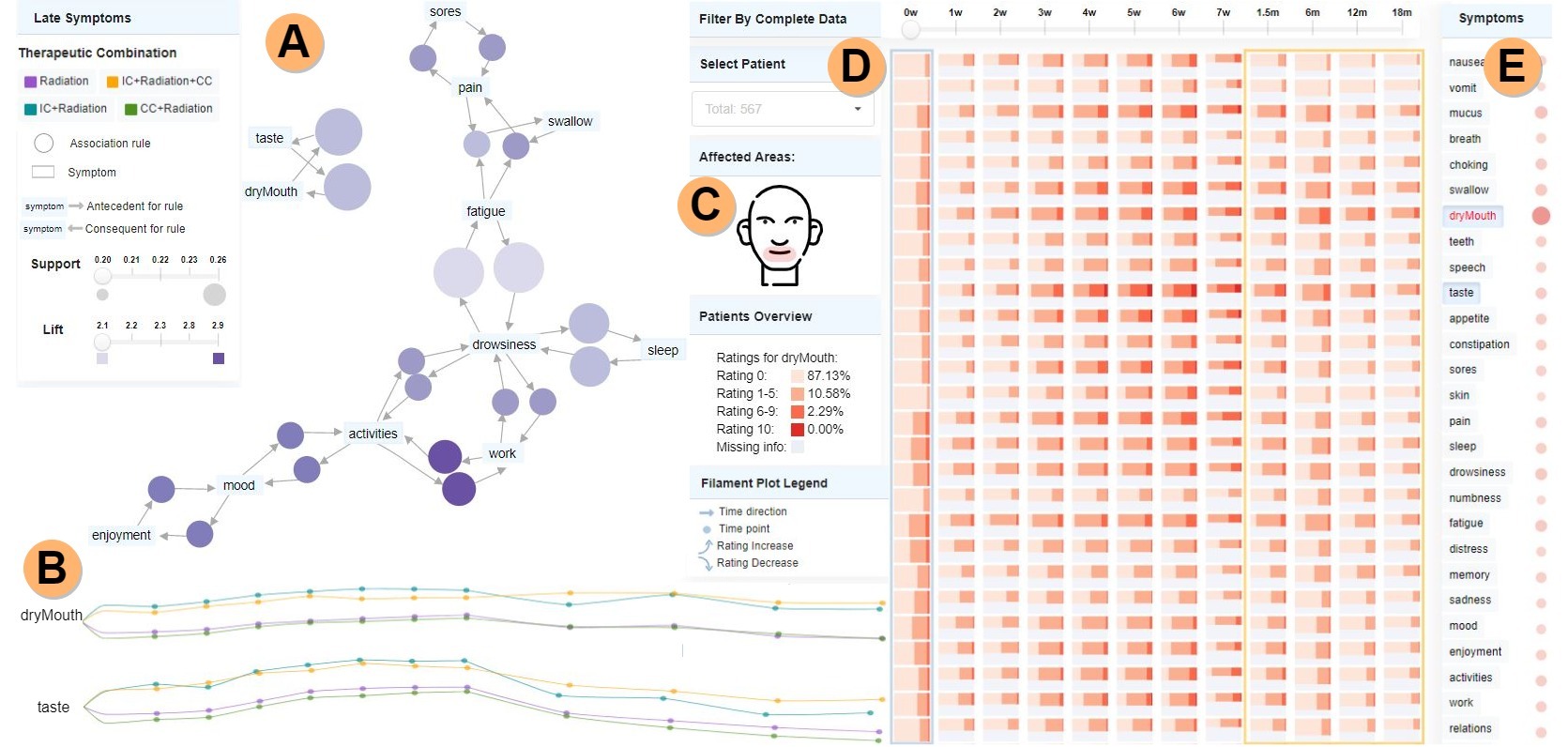}
  \caption{THALIS analysis of longitudinal symptom data. A) Association Rule Diagram panel, showing here association-rule-mining (ARM) relationships among the most frequent late-stage symptoms; rules are represented using bubbles, with size and color encoding the support and lift metrics. B) Symptom trajectory panel---filament plots encode here mean rating values per therapeutic combination, with more frequent observations in the acute stage (left-end) than in the late stage (right-end). C) Sketch of areas affected by the selected symptoms (dry mouth and taste). D) Cohort symptom panel showing via summarization with shade and height the percentile rating distribution. E) Correlation matrix showing associations with the selected symptom. Image cropped and edited for in-print legibility.}
  \label{fig:teaser1}
}

\vgtcinsertpkg
\begin{document}

%% The "\maketitle'' command must be the first command after the
%% "\begin{document}'' command. It prepares and prints the title block.

%% the only exception to this rule is the \firstsection command
\firstsection{Introduction}

\maketitle

% \section{Introduction} %for journal use above \firstsection{..} instead
Thanks to advances in therapeutic care, nowadays cancer patients may survive for years after treatment. However, they are plagued with long-lasting or permanent residual sequelae, whose severity, rate of development, and resolution after treatment vary largely between survivors~\cite{christopherson2019chronic, wentzel2020precision, wentzel2019cohort}. At the same time, patient questionnaires and electronic health records storing such patient responses are leading to larger than ever oncological symptom data collections. These symptom data collected from cohorts of patients~\cite{rosenthal2007measuring} offer important information that can improve clinical decision-making and individual care delivery both during and after treatment~\cite{marai2018precision, sheu2017conditional}, and could be critical for the efficient detection and resolution of longitudinal symptoms. These factors have led to healthcare provider demands to better understand symptom development and prevention based on cohort data. 

However, the meaningful interpretation at the individual patient level of symptom repositories is plagued by data and analysis issues that have prevented their practical use in clinical care. These issues include the wide range of symptoms, their partial co-occurrence, their variability among patients and across time, and, in the case of head and neck cancers (HNC) and other cancers that employ radiation therapy, further symptom dependency on the anatomical location of the tumors and the course of therapy prescribed. To explore these issues, symptom cluster research aims to identify co-occurring symptoms and to understand the underlying mechanisms that drive these clusters, often using machine learning~\cite {miaskowski2017advancing, skerman2009multivariate}. At the same time, HNC analysis results based on factor analysis (e.g., PCA) do not always scale to larger patient datasets~\cite{biggs2021association}. Furthermore, due to methodological limitations, symptom research analyzes either individual symptom evolution or symptom clusters at a single timepoint. Consequently, there is growing interest in alternative machine learning approaches for this type of longitudinal data. Last but not least, these approaches need to make sense in an applied healthcare setting and need to be actionable by clinicians. Therefore, there is also growing interest in mixed human-machine analysis, and a need to leverage and balance computational and human effort for symptom data analysis. 

In this work, we present an interactive data mining environment to support the clustering, exploration, and analysis of longitudinal symptoms collected from cohorts of cancer patients. Our approach intertwines association-rule and factor analysis unsupervised models with custom visual statistical encodings and visual analysis, in order to estimate the longitudinal symptom evolution of an individual patient, in the context of cancer therapies and similar patients. This visual analysis methodology was successfully developed through an interdisciplinary, remote, geographically-distributed collaboration. 

This work contributes: 1) a description of the application domain data and tasks, with an emphasis on the multidisciplinary development of clustering tools for symptom data in cancer therapy; 2) the design of a novel blend of data mining and visual encodings to predict and explain longitudinal symptom development, based on an existing cohort of patients; 3) the description of customized interactive encodings: interactive association-rule diagrams, filaments, and percentile heatmaps; 4) an implementation of this approach in a visual symptom explorer named THALIS: THerapy Analysis of LongItudinal Symptoms (Fig.~\ref{fig:teaser1}); 5) a qualitative evaluation by domain experts using an existing head and neck symptom repository; 6) a start-to-end description of the design process and of the lessons learned from this successful, multi-site remote collaboration.

\section{RELATED WORK}
\noindent\textbf{Electronic Medical Records and Cohort Visualization.} Electronic Medical Records (EMR) store patient longitudinal information, often in the form of time series. In general, time-series visualization has utilized point graphs, circle graphs, line graphs~\cite{harris1999information}, parallel coordinate plots~\cite{inselberg1997multi}, or stacked bar charts and their variations~\cite{visualAnalyticsForTimeData} to encode time-oriented nominal, ordinal or quantitative data, \textcolor{black}{including in cancer~\cite{steinhauer2020comprehensive, muller2019interactive, zebralla2020obtaining}}. For EMR data, Plaisant et al. have introduced personal patient summary visualization using timelines~\cite{plaisant1996lifelines,wang2008aligning, wang2009temporal }, or matrix-based representations~\cite{du2016eventaction}. Loorak et al.~\cite{loorak2015timespan} proposed a stacked bar graph approach to explore patients' treatment processes, while Baumgartl et al.~\cite{baumgartl2020search} explored storyline visualizations from EMR to detect pathogen outbreaks. Rogers et al.~\cite{rogers2019composer} showed outcome trajectories of different patient procedures using line charts. However, most of these approaches are not scalable for large EMR datasets. Wong et al. have employed summarization techniques to overcome issues of scale via tree-based encodings~\cite{wongsuphasawat2011lifeflow} and Sankey-based representations~\cite{wongsuphasawat2011outflow}, while Karpefors's tendril plot~\cite{karpefors2018tendril} introduced a clustered timeline view of outliers and trends for dense clinical trial data. However, none of these approaches include details about individual patients. In contrast, we consider scalable encodings for patient cohort data, and indicate incomplete data and uneven time steps.

%However, none of these approaches include details about individual patients. Furthermore, none of these methods account for the representation of missing data values. In contrast, \textcolor{black}{we} consider scalable encodings for patient cohort data, and indicate incomplete data and uneven \textcolor{black}{time step}s.

In healthcare, patient cohort visual analysis applications span disease evolution statistics extracted from EMRs~\cite{wongsuphasawat2011outflow, huang2015richly}, cohort history comparison~\cite{zhang2014iterative, chui2011visual, bernard2014visual}, inter-cohort medical image attribute comparison~\cite{maries2013grace, steenwijk2010integrated, klemm2014interactive}, survival risk analysis in cancer~\cite{marai2018precision}, and cohort heterogeneous medical data analysis~\cite{turkay2013hypothesis, angelelli2014interactive}. As often the case in clinician-driven visual analysis based on statistics, the visual encodings in these works include conventional representations such as histograms~\cite{baumgartl2020search}, bar charts~\cite{loorak2015timespan}, time-series plots~\cite{gotz2014decisionflow, jin2020carepre}, matrices~\cite{malik2015cohort, du2016eventaction}, radial charts~\cite{guo2019visualizing}, and scatterplots~\cite{kwon2018retainvis}. Similarly, our work builds on patient cohort data, however, our focus is on interpreting individual patient data in the context of similar patients, and contributing visual encodings and workflows which improve the human-machine analysis of symptom data.

\noindent\textbf{Human-Machine Integrated Cohort Cluster Analysis.} Cohort analysis uses unsupervised learning methods such as factor analysis (e.g. PCA), partitional (e.g. K-means), or hierarchical (e.g. agglomerative) clustering. Cluster analysis is traditionally visualized using methods such as scatterplots~\cite{metsalu2015clustvis}, matrices ~\cite{ raidou2018bladder}, radar charts~\cite{marai2018precision}, dendrograms~\cite{eraj2017long}, and heatmaps~\cite{abdullah2020recent}. \textcolor{black}{Temporal clustering is an open problem in symptom research due to the issue of missing data~\cite{madiraju2018deep, ansari2019spatiotemporal, tosado2020clustering}}. Additionally, cancer patient clustering takes into account clinical variables such as the disease stage, treatment plans, medication, treatment toxicity etc.~\cite{ma2017integrate,zdilar2018evaluating}. For HNC patients, Wentzel et al. have introduced spatially-informed distance measures and clustering approaches to group patients based on similarity~\cite{wentzel2020precision, elgohari2019cohort, wentzel2021precision}, although they did not consider symptom data. Gunn et al.~\cite{gunn2013high} and Rosenthal et al.~\cite{rosenthal2014patterns} have studied specifically symptom burden for HNC patients by clustering patients based on reported symptom ratings and clinical covariates to find similarities between symptoms and HNC patients using heatmaps and cluster heatmaps, but do not consider temporal data, nor do they analyze patients that underwent specific treatments, respectively. In contrast, our approach explores groups of similar patients based on symptom load, while also capturing temporal changes in their symptoms. Moreover, we consider the impact of different treatment plans.

Cohort analysis often relies on domain expert interaction to help support human-machine integrated workflows. For more general clustering, several interfaces have afforded user interaction for iterative re-clustering and visualization of unstructured cluster data~\cite{ cavallo2018clustrophile, cava2017clustervis}, although these rely on generic abstract encodings such as scatterplots. Other tools support model building for biostaticians~\cite{dingen2018regressionexplorer}, although these do not consider spatial or temporal outcomes, and are targeted towards statisticians and not clinicians. Angelelli et al.~\cite{angelelli2014interactive} proposed an interactive system for hypothesis generation with retrospective cohort study data using a data-cube-based model that used linked views for spatial and nonspatial data. Other applications have integrated interactive interfaces with application-specific visual encodings with linked views~\cite{wentzel2019cohort, wentzel2020explainable, furmanova2020vapor} to support active collaboration between data analysts and domain experts.  However, none of these approaches consider temporal changes in outcome data or nuanced quality of life outcomes, and do not account for missing data.

\noindent\textbf{Rule Visualization.} Association rules have been visualized via scatterplots, matrix views, node-link representations, mosaic plots, and parallel coordinates plots, as indicated by two surveys~\cite{bruzzese2008visual, jentner2019visualization}, and also as grouped matrices~\cite{hahsler2011visualizing}. More generally, rule-based modeling and visualization are common across domains that seek to understand causality. Colored shapes have been used to indicate information flow in interacting processes~\cite{elmqvist2003growing}. In biological modeling, interactive node-link visual representations have been used for rule-based intracellular biochemistry~\cite{smith2012rulebender, forbes2017dynamic}.  Visual causal vectors have been used to indicate causality between data elements~\cite{ware1999visualizing}, and animated causal overlays have been used to highlight causal flows and to indicate the relative strength of the causal effect~\cite{bartram2008animating}. Whereas our work seeks to identify temporal relationships among data based on association rules, these relationships are not necessarily causal, and they have different features than biochemical pathways.

\section{BACKGROUND}

\noindent\textbf{HNC Therapy and Symptom Collection.} HNC treatment is a complex, longitudinal process that utilizes a variety of therapies, and whose cornerstone is radiotherapy. For example, patients may be prescribed chemotherapy first (induction therapy), and then radiotherapy, or they may be prescribed both chemotherapy and radiotherapy concomitantly. The type of treatment prescribed can result in both short-term (acute, or during treatment) symptoms and in long-term (late, or after treatment) or even permanent sequelae affecting the patient's quality of life.  

In addition to clinical and imaging data~\cite{elhalawani2017matched}, continuous efforts at MD Anderson have included over 1000 patients in a standardized symptom and quality of life monitoring program. The questionnaires are collected on paper at discrete time points, i.e., weekly at the time of the treatment appointment. The questionnaires are based on MDASI (MD Anderson Symptom Inventory)~\cite{Cleeland2000MDASI}, a multisymptom patient-reported outcome measure for clinical and research use. MDASI's thirteen core items include symptoms found to have the highest frequency and/or severity in patients with various cancers and treatment types, whereas the additional MDASI-HN inventory~\cite{rosenthal2007measuring} considers nine symptoms specific to HNC, such as swallowing difficulties, and six additional symptoms that interfere with major activities of daily life, such as enjoyment of life. The compliance rates within head and neck trials are between 60\% and 90\%. However, these patient-generated health data have not been utilized  so far in direct patient care, due to a lack of computational hybrid analytics connecting therapy with the side effects and health state of the patient.

\noindent\textbf{Symptom Clustering Research.}
Cancer patients experience multiple co-occurring symptoms often related to each other and to the therapy applied; however, much of symptom clustering research focuses on single symptoms. In contrast, the term "symptom cluster” (SC) denotes two or more interrelated symptoms that develop together and may or may not be caused by the same underlying mechanism. Several studies have identified symptom clusters in cancer patients~\cite{fan2007symptom, aktas2010symptom, dong2016symptom}, though symptom cluster research is still an emerging field. The two most common methods used to determine SCs are: factor analysis (e.g., principal component analysis, i.e., PCA)~\cite{kim2013analytical, kirkova2011cancer, skerman2009multivariate}  and cluster analysis (e.g., hierarchical agglomerative clustering)~\cite{miaskowski2006subgroups, ferreira2008impact, gwede2008exploring, illi2012association}. However, these approaches have not dealt with changes in symptoms over time, which remains an elusive goal.

Association Rule Mining (ARM), introduced by Agrawal and Srikant in 1994~\cite{agrawal1994fast}, is an alternative unsupervised data mining method, used to identify interesting relationships within data. ARM has been applied to risk management and marketing~\cite{gupta2014survey,kaur2016market}, and more recently, in clinical settings~\cite{kost2012exploring}, although not in symptom clustering.

\section{DESIGN}
\subsection{Collaboration Setting and Design Process}

Our system was developed through a remote collaboration between three different research groups over the course of two years. During this collaboration, our visual computing research group worked closely with oncology and data mining experts. The core team includes 3 radiation oncology experts with clinical and research experience, a senior data mining expert, a data-mining graduate student, and a team of visual computing researchers with varying expertise.  Our team met weekly to produce informative, mixed machine-human analyses of longitudinal symptom data collected from HNC patients who were undergoing treatment at the MD Anderson Cancer Center in Houston, Texas. This work is part of a longer, six-year-long collaboration between the lead investigators who had been working together on a series of related projects using oncology patient data.

Due to the long-term and remote nature of our collaboration, spanned on three sites, we employed team-science principles~\cite{marai2018precision}. Our design process blended an agile design process based on regular team meetings along with an Activity-Centered-Design (ACD) approach to the design of the visualization system~\cite{marai2017activity}. The ACD paradigm is an extension of human-centered-design, with emphasis on user activities and workflow. \textcolor{black}{We note that in the ACD paradigm, the value of a tool depends on the value of the activity, not only on the number of people who use the tool (e.g., a tool serving the two researchers who will find a cure for Alzheimer’s has no lesser value than a tool serving a larger population who are selecting pet names)~\cite{marai2020fabric}. Thus, the ACD paradigm is particularly well suited for tools in scientific research, particularly when we consider the scarcity of trained domain experts, as opposed to the large availability of untrained users, and the importance of slow thinking~\cite{kahneman2011thinking}, including scaffolding.} 

Through a series of iterations, the research team met to define functional specifications, prototype the interface, evaluate prototypes, and decide on changes in the specifications. Moreover, because this approach was designed around developing interfaces that can be shared and designed remotely during the COVID-19 pandemic, our approach proved to be an effective alternative to approaches that rely on in-person group meetings. Additionally, because the ACD paradigm is focused on supporting the collaborators' activities, our collaborators stayed motivated to continue to attend meetings even during circumstances that required remote meetings and exceptional work conditions for clinical practitioners~\cite{norman2013design}.

\subsection{Activity and Task Analysis}
\textcolor{black}{THALIS serves oncologists who have experience in symptom research. Our collaborators also had extensive experience }using basic unsupervised machine learning methods such as factor analysis via principal component analysis (PCA), which they had used to determine that symptom burden varies over time and over patient populations. However, PCA results obtained on smaller datasets did not generalize on larger datasets, so over the course of the project, the group's interests shifted from PCA to alternative approaches. Furthermore, predicting the symptom trajectory of an individual patient in the clinic based on the population data in the repository was not possible computationally because of data issues. Additionally, the oncologists expressed frustration due to repeated patient failures in following instructions aimed to reduce the symptom burden, such as following a prescribed regimen of swallowing exercises or taking the prescribed pain medication. The physicians felt that having the means to explain to patients a predicted symptom trajectory, in the context of other patients, could be beneficial in terms of therapy adherence. 

Accounting for evolving requirements and specifications, we summarize the project activities and their corresponding visual analysis tasks as follows:

\vspace{0.2em}
\textbf{A1.} \textcolor{black}{Analyze} alternative symptom clustering approaches, and apply them to an existing symptom dataset
\begin{itemize}
\itemsep0em \vspace{-0.8em}

\item  T1.1. For each approach, show similar patients, based on symptom severity at a specific time point
\item \vspace{-0.4em} T1.2. For each approach, detect correlations among symptoms, during and after treatment 
\item \vspace{-0.4em} T1.3. For each approach, detect patient outliers and trends
\end{itemize}

\vspace{-0.6em}
\textbf{A2.} Analyze longitudinal symptom progression in the dataset, with particular emphasis on the acute versus late stage of symptoms, and different therapy options
\begin{itemize}
\itemsep0em \vspace{-0.8em}
\item   T2.1. Analyze the patient symptom trajectories as a whole, by therapy type, and by stage
\item  \vspace{-0.4em}  T2.2. Compare symptom trajectories by therapy type
\item  \vspace{-0.4em}  T2.3. Summarize symptom ratings for the entire cohort, by stage 
\end{itemize}

\vspace{-0.6em}
\textbf{A3.} Map an individual patient to its relevant cohort, and explain their longitudinal symptom trajectory in the context of the cohort in an actionable manner
\begin{itemize}
\itemsep0em \vspace{-0.8em}
\item T3.1. Show an individual patient in the context of the cohort
\item \vspace{-0.4em}  T3.2. Display demographic and diagnostic patient data, and indicate patients with similar diagnostic attributes
\item \vspace{-0.4em}  T3.3. Display the anatomical locations affected by a symptom
\item  \vspace{-0.4em} T3.4. Filter a patient's symptoms by association rule
\end{itemize}

%\vspace{-0.5em}
Our evaluation describes example workflows centered on these activities. Non-functional requirements included a request for the A3 data to be displayed in a manner amenable to audiences with low visual literacy, awareness of variability in symptom ratings across patients, and awareness of missing data.

\subsection{Data Analysis}
In accordance with the ACD paradigm for data visualization~\cite{marai2017activity}, the project requirements were based on a starter dataset, which was then expanded during the duration of the project. \textcolor{black}{ Patients who had completed fewer than two questionnaires were not included in the analysis. The final dataset included 699 HNC patients. }

For each patient, two types of information were recorded:  1) Patient demographics and diagnostic data, which covered three attribute types: quantitative data (e.g., age, weight, or the total radiation dose); ordinal data (disease stage), and nominal data (e.g., therapeutic combination); and 2) Longitudinal symptom data, as time-series attributes with quantitative values (ratings for 28 symptoms) over a maximum of 12 time points. The symptoms were further grouped in three categories: core symptoms common for all cancer types (fatigue, disturbed sleep, distress, pain, drowsiness, sadness, memory, numbness, dry mouth, lack of appetite, shortness of breath, nausea, and vomiting), HNC specific symptoms (difficulty swallowing, difficulty speaking, mucus in throat, difficulty tasting food, constipation, teeth/gum issues, mouth/throat sores, choking, and skin pain), and ratings of symptoms' interference with daily life (work, enjoyment, general activity, mood, walking, relationships). The symptoms were rated on a 0-to-10 scale ranging from "not present” (0) to  "as bad as you can imagine” (10) for the core and HNC specific items, and  from "did not interfere” (0) to "interfered completely” (10) for the interference items. \textcolor{black}{Each patient rated all 28 symptoms during a questionnaire completion (time point).}

The dataset included a total of 12 time points. Because of the desired longitudinal aspect of the analysis, we separated these points into three categories: baseline (week 0), acute stage (on-treatment period), and late stage ($>=$ 6 weeks after treatment). For acute time points during treatment, data was collected every week (at most 7 weeks), while after treatment, time points data was collected at lower granularity, at 6-weeks, and 6-, 12-, or 18-months post-treatment. \textcolor{black}{Previous timepoint values were substituted for missing values; missing baseline values (i.e., for the first timepoint) were marked with 0.} Patients with no symptoms recorded during the acute or late phases were not included in the analysis for that time frame. 
%Symptoms with missing scores were replaced with 0s.

\subsection{Environment Design}
The design followed a parallel prototyping approach~\cite{dow2010parallel}, a method proven to lead to better design results by opening up the visual encoding and interaction space, which in turn elicits more detailed and constructive feedback than in serial prototyping. \textcolor{black}{THALIS was implemented in Python and JavaScript with the D3.js library~\cite{bostock2011d3}.} The top design is based on coordinated multiple views of the data, in order to support both layering and separation of information and workflow components, and the ability to integrate visually heterogeneous data. A main clustering panel allows the analysis of patient groupings based on similarity (Fig.~\ref{fig:label2}), respectively the analysis of symptom groups via association rule mining (Fig.~\ref{fig:teaser1}.A). A second main panel supports the longitudinal analysis of patient symptoms (Fig.~\ref{fig:teaser1}.B), in coordination with the other panels. The remaining panels supports explicitly the context-analysis of cohort symptom data. The panels are connected through explicit filtering operations, brushing and linking.

\subsubsection{Clustering Panel}
\begin{figure}[t!]
\includegraphics[width=\linewidth]{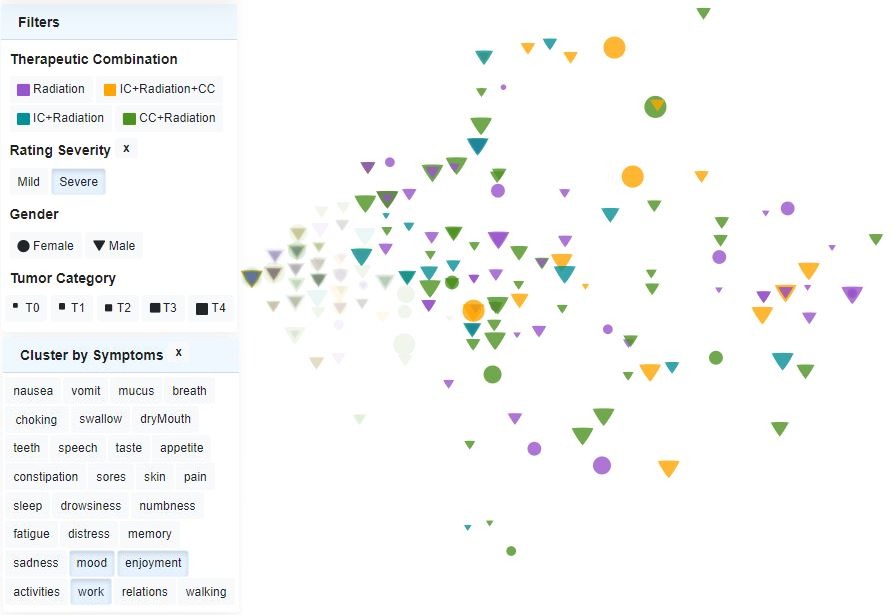}
\caption{Custom scatterplot of patients at a specific time point, for a selected rating severity. Left position is associated with a lower symptom burden, calculated based on the symptoms selected in the list. Shape, size, and color encode demographic, diagnostic, and therapy features (see legend). In this example, highlighted patients correspond to the high rating severity group, indicating that the three symptoms selected  (mood, enjoyment, and walk) severely affect the vast majority of patients across all therapies, genders, and tumor sizes. Outliers are easily noted.}
 \label{fig:label2}
\end{figure}

Because of the experts' interest in activities A1 and A3, the clustering panel shows a therapy cluster view of patients (Fig.~\ref{fig:label2}). Alternatively, the panel shows an association graph view of symptoms (Fig.~\ref{fig:teaser1}.A), illustrating the two main clustering approaches of this project (A1). These views are coupled with computational modules for clustering.\\

\noindent\textbf{Therapy Cluster View.}  In prior research, the clinicians had analyzed a subset of the patient data using factor analysis and had identified distinct groups of patients with high, medium, and low symptom burden, depending on the therapeutic combination, which they had illustrated via heatmaps and dendrograms. However, they were also aware that the heatmap representation did not illustrate well outliers in the patient dataset, nor did it support well individual patient analysis, and they were also not confident about the therapeutic distinction between these groups.  We agreed that a scatterplot view, color-mapped to the different therapies, would serve activities A1 and A3 better, by capturing more clearly individual patients and cohort patterns in the data.

We first organized the symptom ratings into a patient-symptom matrix for the selected time point, where each element $(i,j)$ corresponds to the score given to symptom $j$ by patient $i$ at that time point. Prior research in symptom cluster for HNC~\cite{gunn2013high} had applied hierarchical clustering using Ward’s method~\cite{wardsMethod} with Euclidean distance on the patient-symptom matrix to group patients based on their raw symptom ratings. \textcolor{black}{After alternative clustering with complete and average linkages, we found that Ward’s method generated larger, more informative groups of high symptom patients, which made sense to the clinicians. We identified two patient groups with high and low symptom burden (T1.1).} This two-group clustering was preferred by clinicians, who found it easier to compare two groups instead of more. The axes of the scatterplot correspond to the first two components obtained by applying PCA to the patient-symptom matrix. \textcolor{black}{Clusters for a specific time point are extracted and displayed, while clusters for different timepoints can be investigated via the time slider, which will update the scatterplot.}

The scatterplot was customized to separately capture acute and late symptom burden distribution as identified by the symptom clusters, and to reflect via marker color, shape, and size the therapeutic combination administered to each patient, their gender, and their disease stage (T3.2) (Fig.~\ref{fig:label2}). The data can be filtered by attributes, and filtering operations update the other views. A filtering control panel serves double duty, by also providing the plot legend. This customized scatterplot encoding effectively captured the symptom distribution across the patient population, patient outliers, and therapeutic distribution across the data (T1.1, T1.3, T2.3).

To assess the symptoms’ impact on clustering, we also provide an option for dynamically recalculating the clusters based on user-selected subsets of symptoms (Fig.~\ref{fig:label2}) and update the scatterplot accordingly.\\

\begin{figure*} 
\centering
 \includegraphics[width=0.8\linewidth]{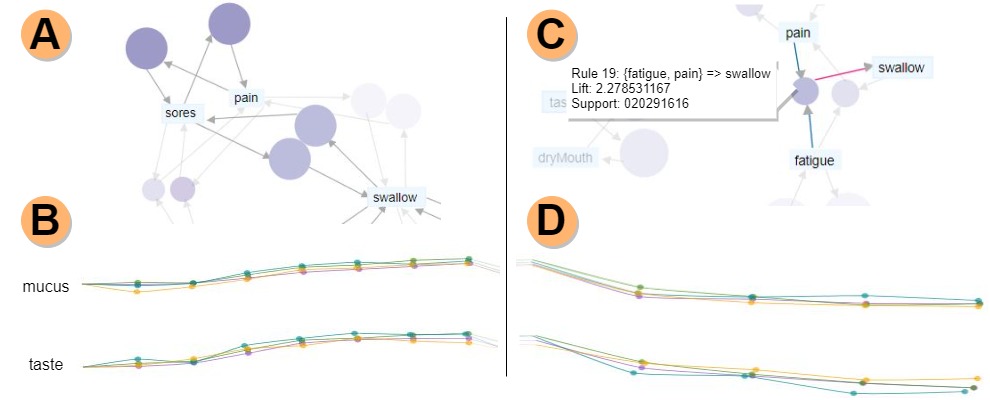}
\caption{Acute vs. late phase analysis. A) Association rule diagram for the acute phase. Rules are filtered based on support (frequency) and lift (dependency between symptoms); other rules are faded in the background. B) Mean rating value filament plots for all therapies, with the acute phase highlighted. All therapies follow similar trajectories for both mucus and taste, and towards the end of the acute phase, taste has a considerable increase in ratings for all therapies. \textcolor{black}{C) Association rule diagram for the late phase, showing the antecedents (fatigue, pain) and consequent (swallow) for rule 14.} D)  Mean rating value filament plots, showing a slightly different trajectory for IC+Radiation.}
\label{fig:label3}
\end{figure*}

\noindent\textbf{Association Rule Diagram View.} Driven by the factor analysis limitations discussed earlier, this project pursued Association Rule Mining (ARM) as an alternative, novel approach to symptom cluster analysis (A1). ARM is an unsupervised data mining technique for identifying relationships within the data~\cite{agrawal1994fast}. In marketing applications, an association rule in the form \textbf{$X \rightarrow Y$} indicates the pattern that if a customer purchases $X$, they will also purchase $Y$, where the patterns are extracted from relational data expressed as transactions. Similar to the strong positive correlations found between items in a supermarket basket, relationships within clinical data can help identify disease comorbidities~\cite{kost2012exploring, huang2008ComorbidSleepApnea, leejin2018comorbidity}. 
%https://www.overleaf.com/project/605b7a24007f8125bdf62955

\begin{table}[th]
\caption{Example of 3 transactions containing \textcolor{black}{4 symptoms: fatigue, drowsiness, pain, and swallow.}}
\label{table:table1}
\begin{minipage}{.5\textwidth}
    \centering
\footnotesize
    \label{symptom_transactions}
    \begin{tabular}{l|llll}
    tid  & items                 &  &  &  \\
    \hline
    001 & fatigue, drowsiness       &  &  &  \\
    002 & pain, drowsiness &  &  &  \\
    003 & fatigue, pain, \color{black}{swallow} & & & \\
    \end{tabular} \\
%    003 & fatigue, pain, \color{blue}{drowsiness} & & & \\
\end{minipage}
\hspace{.3cm}
\end{table}

In this project, we extended the potential of ARM to symptom clustering applications. To this end, we adapted the most common ARM method to our problem: the Apriori algorithm~\cite{agrawal1994fast}, for frequent item-set mining and association rule learning. In our approach, the symptoms experienced at each time point by each patient are treated as a transaction.  The algorithm first identifies frequent symptoms to determine sets of symptoms that co-occur with high certainty and then extends to larger symptom sets. Table~\ref{table:table1} contains an example of three "transactions" from our data. \textcolor{black}{Transactions were extracted from existing questionnaires. Missing ratings for a symptom within a questionnaire implied that the symptom was not included in the transaction. If a patient was missing an entire questionnaire, no transaction was generated for that patient. The ARM was performed using all the available data and no data imputation was performed.}

We followed Agrawal and Srikant's proposed association rule~\cite{agrawal1994fast} in the form:
$$X \rightarrow Y$$
\noindent which indicates that if a patient suffers from symptom $X$ (the antecedent), they will also be affected by symptom $Y$ (the consequent). Based on \textcolor{black}{the first transaction in} Table~\ref{table:table1}, such a rule can be: 
$$\{\text{fatigue}\}\rightarrow \{\text{drowsiness}\}$$
\noindent where $\{$fatigue$\}$ is the rule antecedent and $\{$drowsiness$\}$ is the consequent. \textcolor{black}{For itemsets larger than this pairwise example (e.g., last transaction in Table 1), either the antecedent or the consequent could contain multiple items. }
 
Two standard measures, \emph{support} and  \emph{lift}, are tuned to filter the association rules by a minimum value. \emph{Support} is the measure of how often the transactions contain both $X$ and $Y$\textcolor{black}{, in our case, how frequently sets of symptoms $X$ and $Y$ occur together}. The \emph{support} of a subset of symptoms $S$ is defined by: 

$$\sigma(S) = \frac{|S|}{|T|}$$
\noindent where $|S|$  is the number of transactions that contain all the symptoms in set $S$ and $|T|$ is the total number of transactions in the dataset. In Table~\ref{table:table1}, $\sigma(\{\text{fatigue}, \text{drowsiness}\}) =  \frac{1}{3}$  as both symptoms appear together in $1$ out of $3$ transactions.

\emph{Lift} is the measure of the importance, or strength of the rule, \textcolor{black}{and it shows how more frequently than we'd expect by random chance do $X$ and $Y$ appear together. \emph{Lift}} is defined as:

$$\lambda(X, Y) = \frac{\sigma(X\,\cup\,Y)}{\sigma(X)\times \sigma(Y)}$$

\noindent where $(X \cup Y)$  refers to transactions that contain both $X$ and $Y$. E.g.: 
\begin{equation*}
\resizebox{.99\hsize}{!}{$\lambda(\{\text{fatigue}\}, \{\text{drowsiness}\}) = \frac{\sigma(\{\text{fatigue}, \text{drowsiness}\})}{\sigma(\{\text{fatigue}\}) \times \sigma(\{\text{drowsiness}\})}$}
\end{equation*}

We applied ARM to each of the acute stage and the late stage (T1.2,T3.4), and empirically chose to illustrate the top 20 rules yielded by this approach, because only a small number of rules were of clinical interest. We chose minimum values for the \textit{support} and \textit{lift} metrics that were suitable for frequent and interdependent symptoms.

\begin{figure*}[ht]
\includegraphics[width=\linewidth]{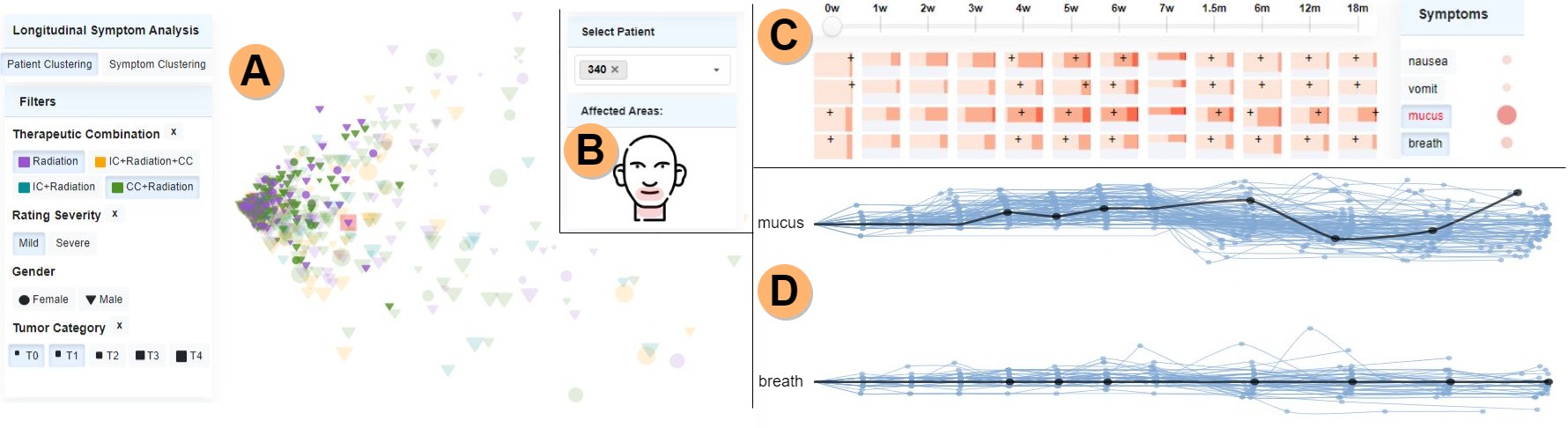}
\caption{Symptom burden analysis. A) Patients in the mild symptom burden cluster, having tumor categories T0 and T1 (current patient, 340, is highlighted in red), with all other patients faded. B) The anatomical sketch shows that the mouth and neck areas are affected by the selected symptoms (mucus, breath) for the current patient. C) The patient's ratings are shown by black marks. In this case, the patient had a low rating for mucus at the first assessment (0 weeks), while at the end of the observation period (18 months post-treatment) the rating increased. D) Filament plots encoding symptom trajectories for the selected symptoms, for the patients filtered in the scatterplot. One filament per patient shows the temporal development for that symptom; black filaments mark the current patient, confirming the mucus rating increase in the late stage.}
\label{fig:label4}
\end{figure*}

From the many possible encodings of ARMs~\cite{hahsler2017arulesviz}, we selected a node-link representation (Fig.~\ref{fig:teaser1}.A), which was deemed by clinicians to be more friendly to broader audiences (A3), and a good fit for the relatively small number of nodes. \textcolor{black}{Graphs are laid out using a force-directed layout algorithm based on statistical multidimensional scaling~\cite{hahsler2011visualizing, north2004drawing}, which results in nodes with high degree being placed centrally. Consistent with this encoding, which is closest to humans arranging nodes manually~\cite{van2008perceptual} or when locating connected clusters~\cite{pohl2009comparing}, the layout is fixed. Other layouts have been tested: the dot layout output a tree-like representation, deemed less desirable, whereas the Distributed Recursive Layout and the Fruchterman-Reingold Layout~\cite{csardi2006igraph} resulted in cluttered diagrams.} We followed established design principles for network visualization~\cite{marai2019ten}: circles encode rules, with larger size and deeper shade denoting higher rule support and lift, respectively, whereas rectangles encode symptoms. \textcolor{black}{Incoming edges for a node indicate which item(s) appear in the antecedent of an association rule, whereas outgoing arrows indicate item(s) in the consequent.  Because the rule directionality is meaningful, rules containing the same sets of symptoms are treated as separate nodes in the graph.} Clicking on a rule highlights the antecedents and consequents of the rule, whereas clicking on a symptom highlights the rules containing that symptom and all the other symptoms in those rules (Fig.~\ref{fig:label3}.C). Rules can be further filtered out based on \textit{support} and \textit{lift} levels (Fig.~\ref{fig:label3}.A).
%Arrows pointing towards a circle mean that the associated symptom is an antecedent (left-hand side) for the association rule. If the arrow points towards a symptom, that symptom is the consequent (right-hand side) for the association rule.

\subsubsection{Symptom Trajectory Panel}
Designing an appropriate encoding for the symptom longitudinal data (A2) turned out to be particularly challenging, primarily due to the nature and richness of the temporal data, the acknowledged variability in ratings across patients, and the missing or uneven time points, which were expected in this context. The design process explored a wide range of possible temporal encodings, many of which suffered from scalability issues and, after several sessions, focused on a promising encoding called a "tendril plot"~\cite{karpefors2018tendril}. A tendril plot is a visual summary of the incidence, significance, and temporal aspects of adverse events in clinical trials, in which individual temporal threads, one per each patient, emanate from a common root and shoot upwards and curl either to the left or to the right depending on whether the next event in the timeline was adverse or an improvement. For clinical trial data, tendrils were shown to create beautiful, compact, naturally clustering pathlines illustrating the positive or negative evolution of each group of patients. The clinicians had also seen this representation and thought it could work (T1.2, T2.2). Whereas promising on paper, unfortunately, the tendril implementation did not yield similarly clean illustrations for the symptom data, because of the much smaller number of time points, the variability in therapeutic sequences, and the variability in patient outcomes, which are not typical of clinical trials. 
  
Numerous design variations yielded a new custom temporal encoding, which we call a filament plot (Fig.~\ref{fig:label4}.D). Filament plots also emanate from a common root, then proceed in a left-to-right direction aligned with the time sequence. Wider timesteps, typical for late stage, are accordingly more widely spaced. Each filament represents the full observation period for a specific patient, with dots along the filament to indicate time stamps. To account for inter-patient rating variability, the curvature degree for the filament at each time step encodes the relative change from the previous rating, where upward rotation indicates worsening symptoms (rating increase), and downward rotation shows symptom amelioration (rating decrease). 

To calculate the rotation, if patient $p$ is located at position $(x_t,y_t)$ at timestep $t$ for a symptom with rating $r$, we compute the next position $(x_{t+1}, y_{t+1})$ at timestep $t+1$ by first calculating the horizontal rotation angle as: 
$$\theta = \frac{\theta_{max} \cdot \Delta r_{t+1}}{2 \cdot \Delta r_{max}}$$
where $\theta {max}$ is the total maximum rotation allowed, whose value is set to $\frac{3\pi}{4}$; \noindent $\Delta r_{t+1}$ is the rating difference between $t+1$ and $t$:
$$\Delta r_{t+1} = r(t+1) - r(t);$$ 
\noindent and $\Delta r_{max}$ is the maximum difference between two rating values which is $10$ in our case. Negative differences between ratings (i.e., rating decreases) produce negative angle values for $\theta$.

Next, we want to rotate $\theta$ degrees relative to the horizontal line $P_1P_2$ defined by the points $P_1 = (x_t,y_t)$ and $P_2 = (x_{t} +l, y_t)$ where $l$ quantifies the time elapsed between $t+1$ and $t$. A higher $l$ indicates that more time passed between $t+1$ and $t$ (i.e., late vs acute). Finally, we rotate $P_2$ around $P_1$ by $\theta$ degrees.

For missing data during the observation period, the associated points are not represented, and we consider no rating change from the previous time points; the surveillance period is represented on each filament until the last recorded time point for each patient. We account for the time ratio between the acute (1 week) and late (months) stages, so the distances illustrated for the acute time points are smaller as opposed to the late time points. Hovering over a filament greys out all the other filaments in the plot.  This interaction helps in the comparison of symptom trajectories for the same patient, and via brushing and linking with the other views, in highlighting the additional patient data (T3.1). 

This compact representation helps in the analysis of symptom evolution trends, by clearly indicating the overall symptom burden (low/high). The representation also helps in spotting outlier trajectories that should be further evaluated and facilitates the discovery of steady vs. variable progression of symptoms. The panel includes two such filament plots, supporting the side-by-side comparison of different symptoms for selected patient groups. To further enhance visual support, during the evaluation of the acute period in the entire THALIS environment, the acute time periods are highlighted in the filament plots, and vice versa for the late period (Fig.~\ref{fig:label3}.B and Fig.~\ref{fig:label3}.D).

In order to better support activities A1 and A2, an additional option uses the same filament encoding, this time with the color mapped to the therapy type, to capture the mean trajectory per each therapeutic combination (Fig.~\ref{fig:teaser1}.B). Since in the therapy case the symptom mean ratings across the population bear meaning, the filaments are spread out according to the mean ratings per therapy (T2.1). This therapy-analysis option helps estimate what treatment plans are less symptomatic, or on the contrary, conduct to high symptom burden.  In addition, to satisfy activity A3, the current patient's filament is highlighted in black in each plot (Fig.~\ref{fig:label4}.D). \textcolor{black}{Whereas reliable automated symptom prediction is an unsolved problem in symptom research, THALIS supports human-machine analysis via trajectory views of similar patients.}

\begin{figure*}[ht]
\includegraphics[width=\linewidth]{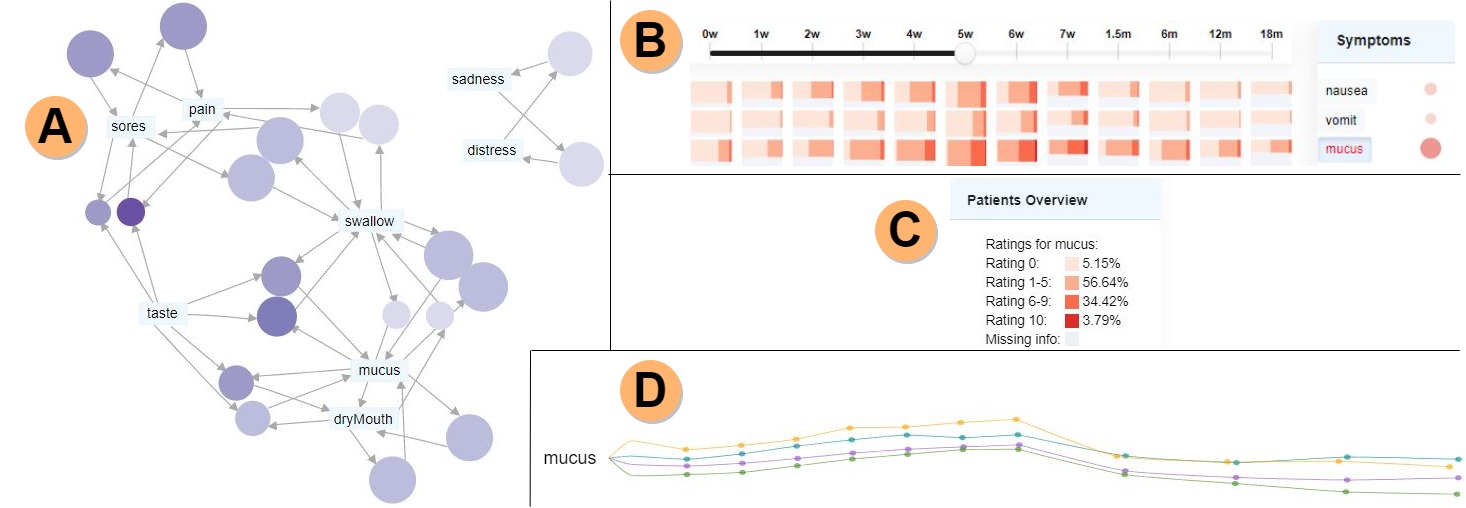}
\caption{Symptom cluster diversity analysis. A) Symptom association graph for the acute phase showing mucus and swallow correlate with many symptoms. Note that the network layout is fixed, and that by construction it places centrally nodes with high degree. B) The percentile heatmap shows a spread of high ratings for mucus along the whole observation period. C) Summary panel for mucus showing that among patients who reported ratings for week 5 during treatment, more than 95\% noted mucus as a present symptom. D) Mean rating filament plot emphasizing rising ratings at the end of the acute phase, especially for the IC+Radiation+CC treatment.
}
\label{fig:label5}
\end{figure*}

\subsubsection{Cohort Symptom Panel}
The last panel explicitly supports activities A1 and A3, and provides an abstract summary of the entire temporal symptom data. As in other fields~\cite{luciani2018details}, and as indicated by our activity analysis, this summary provides context for a specific datapoint, but does not lead the investigation. The panel comprises a percentile heatmap, a correlation matrix, and an anatomical sketch (Fig.~\ref{fig:teaser1}).

The percentile heatmap (Fig.~\ref{fig:teaser1}.D) is a custom representation showing the rating distribution of individual symptoms over time, for the entire patient cohort (T2.3). We arrived at this representation after exploring a variety of alternatives such as stacked line plots, parallel coordinates plots, and radar charts, guided by feedback from collaborators. We settled on a matrix-based layout due to its compactness and to its ability to support small multiple plots. Each row corresponds to a symptom, with rows grouped by symptom category, and each column corresponds to a time point. Each cell in this matrix is a horizontal bar graph showing via shade the percentage of patients reporting within a specific range (0, 1-5, 6-9, or 10) for that symptom, at that time point. The bar height maps the percentage of individuals from the entire cohort who reported the symptom ratings at that time point. The current patient is indicated in this heatmap by cross markers (Fig.~\ref{fig:label4}.C) (T3.1). This encoding proved to be an intuitive way of showing what symptoms produce a higher burden for patients, and when, as well as to indicate how many patients were affected by these symptoms from the entire cohort (T1.2, T2.3).

To support exploration driven by a specific patient (A3), a dropdown selection box is also provided  (Fig.~\ref{fig:label4}.B). A selection in this box highlights the patient data across panels (Fig.~\ref{fig:label4}). A timeline selector further allows the selection of a particular time point in the data (Fig.~\ref{fig:label4}.C), and further interface elements allow selection and analysis of sets of similar patients. Additionally, a compact correlation matrix  (Fig.~\ref{fig:teaser1}.E), along with the percentile heatmap, supports T1.2, by showing the strength of the correlation between a selected symptom and all other symptoms, with circles encoding Spearman's coefficient via color and size. Finally, because a discussion of task T3.3 revealed that patients tend to point to the location of their symptoms, an anatomical sketch (Fig.~\ref{fig:teaser1}.C) supports visual anchoring based on anatomy. Regions in the head and neck affected by the selected symptoms are highlighted in this sketch.

\section{EVALUATION AND RESULTS}
\textcolor{black}{Because no design approach is failproof, although ACD has higher success rates than HCD (63\% compared to 25\%)~\cite{marai2017activity}}, we evaluated THALIS through a combination of multiple demonstrations and case studies involving domain experts, namely a senior data mining specialist and three senior clinical radiation oncology experts. \textcolor{black}{Whereas we recognize these experts as co-authors, not all of them were involved in the development process at all stages.} Two case studies were completed during separate, dedicated sessions, in addition to regular feedback sessions. Because the designers and evaluators were in different locations, and due to COVID-19 constraints, these sessions were conducted remotely using screen sharing and note-taking. The oncology experts directed the exploration using the think-aloud method, while the first author was driving the interface according to their instructions. Both case studies analyze a set of 699 HNC patients, which was significantly larger than prior clinician analyses, and span all activities, A1-A3. Qualitative feedback was also provided during weekly design-driven sessions and was used to improve the overall design of THALIS.

\begin{figure*}[ht]
  \includegraphics[width=\linewidth]{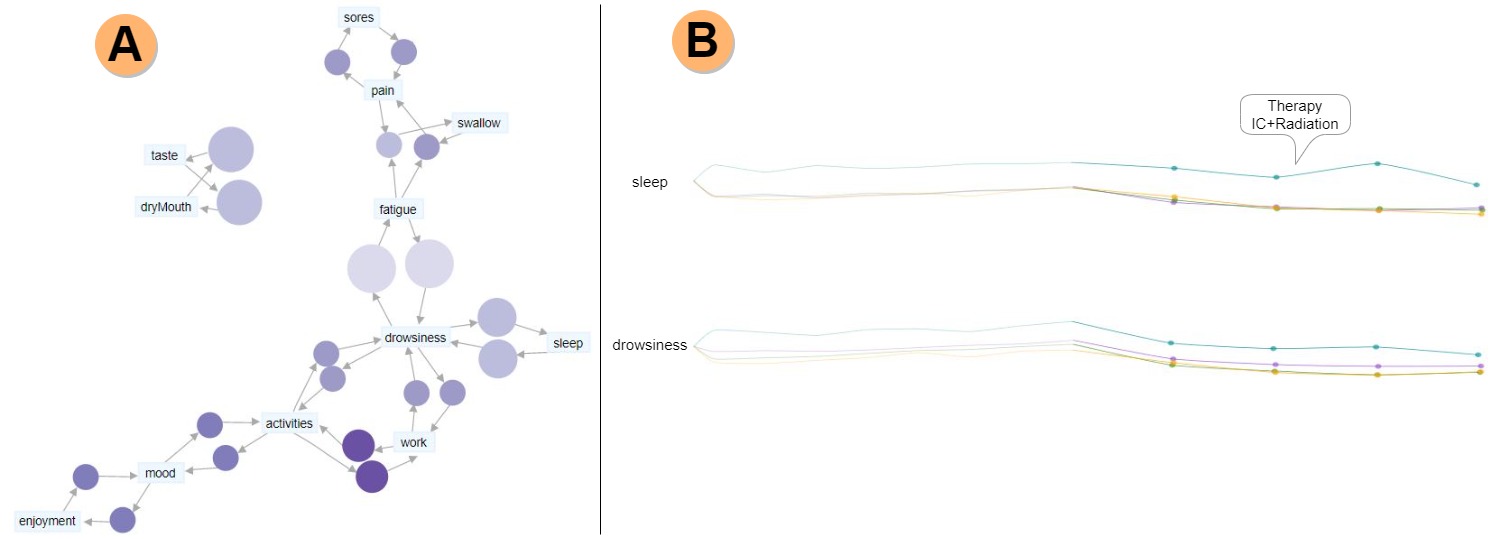}

\caption{Late symptom cluster analysis. A) Symptom association graph showing drowsiness as a central symptom for the late phase. The connection between sleep and drowsiness is expected, as these two symptoms are known to be a factor in  dangerous muscle-mass loss. B) The filament plots show mean rating values, with the late phase highlighted, and the acute phase faded. Notably, in the case of sleep and drowsiness, IC+Radiation is the therapy associated with higher symptom ratings, and it is noticeably different from the other therapy plans.
}
\label{fig:label6}
\end{figure*}

\subsection{Case I: Symptom-Burden Analysis in Radiotherapy}
The study seeked to assess the impact of therapy on symptom burden on this set, and took place before our development of the associative rule model. The oncologists were originally hoping to replicate published analysis results obtained on significantly smaller cohorts of 80 to 270 patients~\cite{rosenthal2014patterns, eraj2017long, kamal2019modeling}. Using the system over the course of several sessions showed, however, that those clustering results were not generalizable to the larger cohort, and so the investigation shifted focus to discovering and analyzing outliers in terms of patient characteristics and symptom trajectories. The study workflow started directly with the therapy scatterplot panel (Fig.~\ref{fig:label4}.A) (T1.1, T3.2). At first glance, most patients were visibly grouped in the left-center part of the plot, suggesting strong similarity. Filtering the patients (T3.1) based on their rating severity revealed that this group corresponded to a mild-rating severity cluster. Further filtering by therapy and tumor category, the experts noted that most of these patients were treated with radiation with or without concurrent chemotherapy (CC) and, not surprisingly, presented a small tumor size and a low symptom burden at the end of the observation period. They concluded that for this group, the therapy plan did not effectively impact the quality of life. Next, the oncologists examined whether a smaller set of symptoms, as in their prior studies, would correlate with patient groupings (T1.1, T1.3). To this end, they filtered data by daily interference symptoms, including, for example, $\{$mood, enjoyment, and work$\}$ (Fig.~\ref{fig:label2}). This time, they found that almost a third of the \textcolor{black}{patients} suffered from high symptom burden in this symptom group. 

Encouraged by this finding, the analysis moved swiftly to the filament plots (Fig.~\ref{fig:label4}.D), to examine the symptom trajectories (T2.2). The plots captured a general trend in most symptom trajectories, namely, a rating decrease post-treatment, with the exception of $\{$numbness, memory, breath$\}$. Moreover, these three symptoms, along with nausea and vomit, exhibited a steady symptom development, with fewer patient outliers or drastic rating changes over time (T1.2). There was, in fact, no correlation between the temporal outliers in the filament plots and the therapy scatterplot outliers. This finding indicated that patients experienced steady ratings for these five symptoms over time, regardless of overall symptom burden or therapy treatment. This observation was of notable interest, and so the analysis moved to examine the cohort context (T2.3). Using the percentile heatmap (Fig.~\ref{fig:teaser1}.D) and the correlation matrix, our collaborators noted that groups of symptoms such as $\{$swallow and dry mouth$\}$, or $\{$taste, appetite, constipation, and sores$\}$ showed higher ratings over time, suggesting possible interrelationship or causative factors between these symptoms. For example, when selecting dry mouth, the panel indicated strong correlations between dry mouth and $\{$mucus, choking, and swallow$\}$, but also with $\{$taste, drowsiness, and fatigue$\}$ as well. Finally, the anatomical sketch layout (Fig.~\ref{fig:teaser1}.C) emphasized which head and neck locations are affected by the selected symptoms (T3.3). In this case, we noted that both dry mouth and taste affected the mouth area. The oncologists are planning studies to verify this set of symptom cluster hypotheses.

\subsection{Case II: Symptom Cluster Diversity}
This study aimed primarily to explore the value of associative rule mining in longitudinal symptom analysis (T1.2). Examining the association diagrams, the oncologists were stunned to find surprising symptom clusters during and post-treatment; in particular, 8 common symptoms for the acute stage (Fig.~\ref{fig:label5}.A), with two strongly coupled subgroups: $\{$distress, sadness$\}$, and $\{$swallow, pain, sores, taste, mucus$\}$; and respectively, 12 frequent symptoms during the late part of the treatment (Fig.~\ref{fig:teaser1}.A), showing symptom clusters such as $\{$taste, dry mouth$\}$, and $\{$sores, pain$\}$. The  experts were impressed to see that the $\{$sores, pain$\}$ cluster is strongly associated with $\{$taste$\}$ in the acute phase, while in the late phase, there is a connection between $\{$drowsiness, sleep$\}$ (Fig.~\ref{fig:label6}.A), which is known to be a factor in  dangerous muscle-mass loss. The $\{$taste, dry mouth$\}$ cluster in the late phase supported our collaborators' previous findings. However, the connection between $\{$fatigue, drowsiness$\}$ in the late phase and the centrality of $\{$mucus$\}$ (Fig.~\ref{fig:label5}.A), as well as the $\{$taste, sores$\}$ connection within the acute graph was unexpected. \textit{"In our group, we have established this arc from taste to dry mouth in late stage, but we haven't thought of the taste to sores link in the acute phase. That is striking."}

The ability to highlight a particular symptom (T3.4) or rule and to filter the rules based on their support and lift (Fig.~\ref{fig:label3}.A, Fig.~\ref{fig:label3}.C) were found essential during the exploration, by helping our collaborators to figure out which symptoms were more persistent or more dependent on each other. For instance, $\{$fatigue, drowsiness$\}$ were the most common symptoms (based on their support) and $\{$activities, work$\}$ the most dependent on each other (based on their lift) in the late phase (Fig.~\ref{fig:teaser1}.A). Insights observed from the symptoms association graphs were further extended using the percentile heatmap (Fig.~\ref{fig:teaser1}.D), revealing the spread of high ratings for $\{$taste$\}$ and $\{$fatigue$\}$ over the whole patient supervision period (T2.3). Moreover, because $\{$mucus$\}$ was usually perceived as an acute symptom, the experts found it remarkable that a large number of patients experienced $\{$mucus$\}$ during the late period as well (Fig.~\ref{fig:label6}). The mean value filament plots
were used to show the mean ratings per time point for each therapy while highlighting the treatment phase of interest (acute/late) (Fig.~\ref{fig:label3}.B, Fig.~\ref{fig:label3}.D) (T2.1). The plots showed that the trends were remarkably conserved over time between therapies, even though their magnitudes might differ. To achieve a better understanding, the option to separate the filaments based on the starting mean rating (baseline) was used (Fig.~\ref{fig:teaser1}.B, Fig.~\ref{fig:label5}.D) which showed a difference in the symptom burden between therapies for the association-identified symptom groups. For example, in the case of $\{$taste$\}$ and $\{$mucus$\}$, in both acute and late phases, the highest rated treatments were IC+Radiation+CC (induced chemotherapy, radiation, and concurrent chemotherapy) and IC+Radiation, while CC+Radiation and Radiation alone were rated lower. Noticeably, in the case of $\{$drowsiness, sleep$\}$, IC+Radiation was remarkably separated from the other treatment plans (Fig.~\ref{fig:label6}.B). The oncologists concluded this case study and the associative approach were a gold mine for their symptom research, by highlighting the diversity of symptom clusters over time.
% i fixed the space problem. you dont need to remove words.. check line 150 that's the command that removes unwanted spaces between paragraphs
\subsection{Expert Feedback}
Because THALIS used participatory design, feedback from the domain experts is implicitly reflected in the final design choices we report. Here, we focus instead on expert feedback related to the current system. The current version of THALIS yielded excellent feedback from the oncology team, often indicating a shift in thinking about their work. We report sample feedback, in relation to our activity analysis A1-A3: 

(A1, A3) Quote from the most senior oncologist: \textit{I gotta be honest, every time I meet with you guys and we see these visualizations, I get so much material for future research. In general, to be fair, my focus in clinical practice [and in helping patients] tends to be on dry mouth and swallowing. I say "We're going to talk about dry mouth and swallowing, cause these two are really bad", and "then there's all the other stuff". And then I see this [the ARM and heatmap and filaments], and here's this other stuff, that is usually at my periphery, but I don't focus on, although patients do mention it.  If I were sitting with a patient and I'd look at this interface and ARMs---I get it, hey, there's actually a LOT of moving parts here [beyond dry mouth and swallowing], and they're related, and they have different time sources. It's sobering.} 

(A1, A2) Both case studies had the team exclaim, on multiple occasions, about being \textit{"blown away", "surprised by that", "that [symptom] spread over time just jumps out at you", "This entire ARM approach is so different [from the approach we've followed in our past research on symptom clusters]. I want to stick a flag in the ground with the ARM work, and look at dose to organs and use ARM to see dose-to-swallowing correlation, based on this spatial structure underneath", "This interface and the ARM provide great preliminary data for so many grants [projects] right off the bat!", "Really impressed", and "[This relationship] is not intuitive, so it's very interesting. And I wouldn't have thought about it. But now, it makes perfect sense. Duh!", "The [filament view] is such a great asset for the interface."}

(A3) The clinician oncologists: \textit{"[THALIS's] ability to go from patient to population is fantastic, I really love it, it's exactly what I need", "I like that when a patient is with [CDF], they want percentages, e.g., 66\% of patients have normal appetite after 12 months, and [THALIS] shows that", "When I see a patient, this [taste-dry mouth] association in the late phase is the default picture I have in my mind. But here I see that also fatigue connects to drowsiness, and that these symptoms show up in the acute phase as well, and that I really need to discuss these issues with my patients." "I can share [this view] with my patients,  to explain that pain and swallowing and fatigue are really tightly related---we don't know if it's causation, but they definitely show up together, so could you please, please, take your pain and  anti-inflammatory meds, and could you please do the swallowing exercises we've talked about?"}

\section{DISCUSSION}
The case studies and the domain expert feedback demonstrate THALIS's value in bridging the gap between machine and human analysis, and its ability to help generate novel insights. Our integrated approach is able to capture longitudinal differences between acute and late stages, while detecting outliers and trends in the symptom and therapy data. More importantly, our approach supports individual patient analysis, while handling a large cohort both computationally and visually. Through an ACD approach, and as indicated by the expert feedback, THALIS successfully serves the core interests of its audience. In conjunction with the clustering panel, the symptom association rule view, the filament plots, and the cohort symptom panel enabled discovering interesting relationships in the data, and in several cases lead to unexpected but insightful results. Furthermore, THALIS couples multiple customized novel visual-encodings with symptom clustering algorithms in the background, enabling the domain experts to explore multiple scenarios and test their hypotheses in real-time. Its use of a multi-view paradigm supports flexible analytical workflows that leverage computational power and human expert knowledge. 

Through close collaboration with domain experts, our solution introduces compact, customized visual encodings for the symptom data: a filament encoding and a percentile heatmap. The percentile heatmap scales well with the number of subjects, by design, at the cost of summarization. \textcolor{black}{Whereas the inherent scalability of filaments with the number of items shown is limited, these encodings successfully abstract the cohort data with the help of similarity-based filtering operations, which are appropriate in this context; for hundreds of dense observations, as common in other problems, tendrils~\cite{karpefors2018tendril} offer a better solution. In further terms of scalability, the ARM graph can provide rules for any number of time points in the late and acute time periods. Still, the graph representation for association rules is suited for a smaller number of rules (less than 100~\cite{hahsler2011visualizing}). The scatterplot and correlation matrix are time point specific, so any number of plots could be generated. On the other hand, some views are prone to clutter. Some of these encodings may have limited generalizability beyond this application domain. In the case of filaments, they work in this application because there is a significant correlation between similar patients' trajectories and because our application emphasizes relative trajectory changes as opposed to absolute values. This type of correlation and relativity may not be true across application domains. However, our custom encodings can be repurposed for other longitudinal problems that feature missing data, as in astronomy or biology~\cite{luciani2014large, hanula2019darksky, smith2012rulebender, ma2017prodigen}. Future work includes longitudinal clustering, applying the ARM approach on sequential data, and interactively changing ARM metrics and the number of rules.}

Reflecting upon this successful design experience, we extract three main lessons for designers dealing with similar problems:

\textbf{L1.} \textit{Use an activity-centered design process, in particular in remote collaborations.} In our experience, following a design process focused on activities as opposed to humans, from requirements engineering to the evaluation against these activities, allowed us to align this project with the core client interests. Because of this alignment with their core activities, we had significant buy-in from clients, providing us with the ability to stay on task and make steady progress, propelled by activity-relevant insights in several meetings. The approach furthermore resulted in a successful remote collaboration, and an eagerness to adopt THALIS in the clinic.

\textbf{L2.} \textit{Use visual scaffolding to introduce custom, novel visual encodings.} Through many design iterations, our solution converged towards custom encodings, such as filaments and the percentile heatmap. These compact encodings are scalable (in the case of filaments, through filtering), and effectively serve the original design aims. We were able to introduce these encodings through visual scaffolding~\cite{marai2015visual} over many meetings with domain experts: small, gradual changes from one iteration to the next. From the other end of the spectrum, established encodings such as scatterplots and node-link diagrams have greater adoption chances in low visual literacy environments. A mix of novel and standard encodings, when following visualization design principles, may facilitate encoding adoption.

\textbf{L3.} \textit{In XAI (explainable AI), emphasize domain sense and actionability.} In healthcare applications like THALIS, that blend visual encodings with alternative AI methodology such as ARM, we found that transparency in the AI model~\cite{lipton2016mythos} was not enough to make the model trustworthy. Beyond transparency, our model outputs, in their node-link representation, made sense to the domain experts because THALIS did not contradict their clinic knowledge, and although it did not confirm earlier findings on smaller cohorts, the experts appeared to gain trust in it. At the same time, it was essential to make this AI model actionable: cohort-based analyses are useful in symptom research, but in the clinic, the emphasis is on the individual patient, their therapy, and their likely symptom trajectory. Building an integrated machine-human system that explicitly supports the need to act on the patient's care served our project well.

\section{CONCLUSION}

In this work we described the activity-centered design of THALIS, a novel environment to support the integrated human-machine analysis of longitudinal symptom clusters as a function of cancer therapy. We described the application domain data and activities with an emphasis on the multidisciplinary development of clustering tools for symptom data in cancer therapy. We also introduced a novel blend of data mining and visual encodings to predict and explain longitudinal symptom development based on an existing cohort of patients and described customized interactive encodings: interactive association-rule graphs, filaments, and percentile heatmaps. The evaluation of the resulting mixed workflows and encodings over an existing head and neck cancer symptom repository with domain experts proves the value of this integrated approach for both symptom research and work in the clinic. Last but not least, we summarized the design lessons learned from this successful, multi-site, remote collaboration. We hope these lessons will help other designers who tackle similar design problems and challenges in human-machine integrated visual analysis.

%% if specified like this the section will be committed in review mode
\acknowledgments{The authors are partially supported by the U.S. National Institutes of Health, through awards NIH NCI-R01CA258827, NIH NCI-R01CA214825, and NIH NCI-R01CA2251, and by the US National Science Foundation, through awards NSF-IIS-2031095, NSF-CDSE-1854815 and NSF-CNS-1828265. We thank all members of the Electronic Visualization Laboratory, and all members of the MD Anderson Head and Neck Collaborative Group.} 
% \cite{christopherson2019chronic}
% \cite{wentzel2020precision}  \cite{wentzel2019cohort} \cite{rosenthal2007measuring}
% \cite{marai2018precision} \cite{sheu2017conditional} \cite{miaskowski2017advancing} \cite{skerman2009multivariate} \cite{plaisant1996lifelines} \cite{wongsuphasawat2011lifeflow} \cite{wongsuphasawat2011outflow} 
% % \cite{Nielson:1991:TAD} \cite{Wyvill:1986:DSS}
% \cite{wang2009temporal} \cite{malik2015cohort} \cite{huang2015richly} \cite{zhang2014iterative} \cite{chui2011visual} \cite{bernard2014visual} \cite{maries2013grace} \cite{steenwijk2010integrated} \cite{klemm2014interactive} \cite{turkay2013hypothesis} \cite{angelelli2014interactive}  \cite{baumgartl2020search}

% $$\frac{3\pi}{4}$$

% \cite{demiralp2017clustrophile}

\nocite{*}
\bibliographystyle{abbrv-doi}

\bibliography{template}

\end{document}